\documentclass[sigconf]{acmart}

\newcommand\footnotetextcopyrightpermission[1]{} 

\makeatletter
\def\@copyrightspace{\relax}
\makeatother

\usepackage{fancyhdr}
\usepackage[utf8]{inputenc}
\title{ZeRO-Infinity: Breaking the GPU Memory Wall \\ for Extreme Scale Deep Learning}
\author{Samyam Rajbhandari, Olatunji Ruwase, Jeff Rasley, Shaden Smith, Yuxiong He \\ \tt{\small \{samyamr, olruwase, jerasley, shsmit, yuxhe\}@microsoft.com} \vspace{10mm} }

\date{}
\usepackage{xspace}
\usepackage[symbol]{footmisc}
\usepackage{graphicx}
\usepackage{multirow}
\usepackage{hhline}
\usepackage{comment}
\usepackage{adjustbox}
\usepackage{xcolor}
\usepackage{pifont}
\usepackage{hyperref}
\usepackage[capitalize]{cleveref}
\usepackage{subcaption}
\usepackage{enumitem}
\usepackage{float}
\usepackage{appendix}
\crefname{section}{Sec.}{Secs.}

\usepackage[switch,columnwise]{lineno} 

\newcommand{\name}{ZeRO-Infinity\xspace}
\newcommand{\zoff}{ZeRO-Offload\xspace}

\newcommand{\mycheckmark}{\color{green} \ding{51}}
\newcommand{\mycrossout}{\color{red} \ding{55}}

\newcommand{\todo}[1]{\color{red} TODO: #1}

\begin{document}
\pagenumbering{Roman}
\maketitle

\thispagestyle{fancy}
\lhead{}
\rhead{}
\chead{}
\rfoot{}
\cfoot{}
\renewcommand{\headrulewidth}{0pt}
\renewcommand{\footrulewidth}{0pt}

\renewcommand{\thefootnote}{\fnsymbol{footnote}}
\footnotetext{}
\renewcommand{\thefootnote}{\arabic{footnote}}

\vspace{20mm}
\section*{Abstract}
In the last three years, the largest dense deep learning models have grown over 1000x to reach hundreds of billions of parameters,
while the GPU memory has only grown by 5x (16~GB to 80~GB). Therefore, the growth in model scale has been supported primarily though system innovations that allow large models to fit in the aggregate GPU memory of multiple GPUs. However, we are getting close to the GPU memory wall. It requires 800 NVIDIA V100 GPUs just to fit a trillion parameter model for training, and such clusters are simply out of reach for most data scientists. In addition, training models at that scale requires complex combinations of parallelism techniques that puts a big burden on the data scientists to refactor their model. 

In this paper we present \name, a novel heterogeneous system technology that leverages GPU, CPU, and NVMe memory to allow for unprecedented model scale on limited resources without requiring model code refactoring. At the same time it achieves excellent training throughput and scalability, unencumbered by the limited CPU or NVMe bandwidth. \name can fit models with tens and even hundreds of trillions of parameters for training on current generation GPU clusters. It can be used to fine-tune trillion parameter models on a single NVIDIA DGX-2 node, making large models more accessible. In terms of training throughput and scalability, it sustains over 25 petaflops on 512 NVIDIA V100 GPUs (40\% of peak), while also demonstrating super linear scalability. An open source implementation of \name is available through \href{https://www.deepspeed.ai/}{DeepSpeed}\footnote{DeepSpeed (\url{https://www.deepspeed.ai/}) is a deep learning optimization library designed to make distributed training easy, efficient, and effective. DeepSpeed has been extensively adopted by the DL community.}.
\section{Extended Introduction}
\label{sec:introduction}

Deep learning (DL) has made tremendous advances in recent years, allowing it to  become an integral part of our lives from powering our search engines to our smart home virtual assistants. Increased model size is at the center of these advancements~\cite{DBLP:journals/corr/bert, gpt-2, T5}, and multiple studies have shown that this trend will continue~\cite{brown2020language,kaplan2020scaling}. As a result, there has been significant investment in training huge models.

In the last three years, the largest trained dense model in deep learning has grown over 1000x, from one hundred million parameters (ELMo~\cite{peters2018deep}) to over one hundred \emph{billion} parameters (GPT-3~\cite{brown2020language}). In comparison, the single GPU memory has increased by a meager 5x (16~GB to 80~GB). Therefore, the growth in model size has been made possible mainly through advances in system technology for training large DL models, with parallelism technologies such as model parallelism~\cite{megatronlm}, pipeline parallelism~\cite{DBLP:journals/corr/pipedream,GPipe,narayanan2020memory}, and ZeRO~\cite{zero,zero-offload} creating a path to training larger and more powerful models. 

The current state-of-the-art in large model training technology is three-dimensional parallelism (3D parallelism \cite{3dparallelism,megatrongithub}), which combines model (tensor-slicing) and pipeline parallelism with data parallelism to efficiently scale DL training to trillions of parameters on hundreds or thousands of GPUs. For example, the DeepSpeed implementation of 3D parallelism can scale to over a trillion parameters on 800 NVIDIA V100 GPUs by fully leveraging the aggregate GPU memory of a cluster~\cite{deepspeed3dblog}.

Despite the capabilities of 3D parallelism for large model training, we are now arriving at the GPU memory wall~\cite{gholami2021}. The aggregate GPU memory is simply not large enough to support the growth in model size. Even with the newest NVIDIA A100 GPUs with 80~GB of memory, 3D parallelism requires 320 GPUs just to fit a trillion-parameter model for training, and scaling to a hundred trillion parameter model of the future would require over 6K GPUs even if we assume a 5x increase in GPU memory in the next few years. We can no longer sustain the continuous growth in the model scale with GPU memory as the bottleneck.

The GPU memory wall also limits data scientists from accessing even the large models of today, especially for fine tuning. Large models are first pretrained on large amounts of generic data, and through fine tuning the same model can be specialized for a wide variety of applications. 
\begin{figure}[!tbp]
  \centering
   \includegraphics[width=0.98\columnwidth]{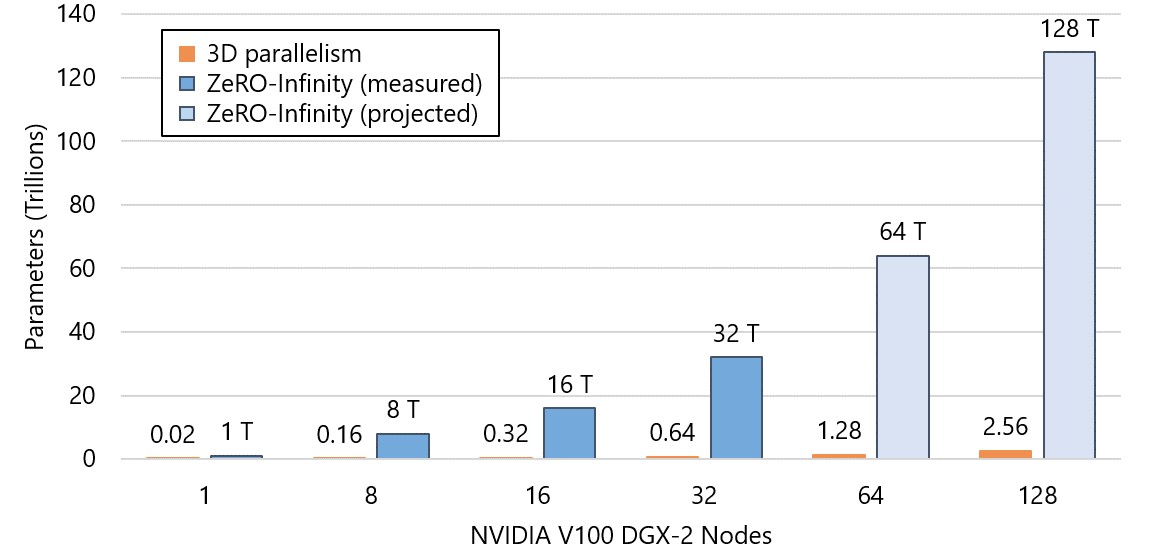}
  \caption{ \name can train a model with 32~trillion parameters on 32~NVIDIA V100 DGX-2 nodes (512~GPUs), 50x larger than 3D parallelism, the existing state-of-the-art.}
\label{fig:max-model-size-128nodes}
\vspace{-10pt}
\end{figure}
While pretraining a model with hundreds of billions of parameters can require millions of GPU compute hours, fine-tuning it is much cheaper, requiring significantly fewer GPU compute hours, and could be done on a single compute node with a handful of GPUs. While such compute resources are accessible to many businesses and users, they are unfortunately restricted by the memory available on these compute nodes, which in turn limits the size of the model that can be fine tuned.  It makes large model fine tuning inaccessible to most researchers and companies that do not have access to massive GPU clusters. For example fine-tuning GPT-3 would require over 8 DGX-2 nodes(128 GPUs) with 3D parallelism to just fit the model for training, even though a single DGX-2 node (16-GPUs) has enough compute to fine-tune it in a reasonable time. 

In addition to the GPU memory wall, state-of-the-art for training massive models is also limited in terms of usability and flexibility. 
As discussed above, 3D parallelism requires combining data, model, and pipeline parallelism in sophisticated ways to get to hundreds of billions or trillions of parameters.
While such a system can be very efficient, it requires data scientists to perform major model code refactoring, replacing single GPU operators with tensor-sliced versions, and splitting the model into load-balanced pipeline stages. This also makes 3D parallelism inflexible in the types of models that it can support. Models with complex dependencies cannot be easily converted into a load-balanced pipeline. 

Given the landscape of large model training, 3 questions arise:
\begin{itemize}[leftmargin=*]
\item Looking ahead, how do we support the next 1000x growth in model size, going from models like GPT-3 with 175 billion parameters to models with hundreds of \emph{trillions} of parameters?

\item	How can we make large models of today accessible to more data scientists who don't have access to hundreds to GPUs?

\item	 Can we make large model training easier by eliminating the need for model refactoring and  multiple forms of parallelism?
\end{itemize}
In this paper, we take a leap forward from 3D parallelism and present \name, a novel system capable of addressing all the aforementioned challenges of large model training. 

\textbf{Unprecedented Model Scale} \name extends the ZeRO family of technology~\cite{zero,zero-offload} with new innovations in heterogeneous memory access called the \textit{infinity offload engine}. This allows \name to support massive model sizes on limited GPU resources by exploiting CPU and NVMe memory simultaneously. In addition, \name also introduces a novel GPU memory optimization technique called \textit{memory-centric tiling} to support extremely large individual layers that would otherwise not fit in GPU memory even one layer at a time. With the infinity offload engine and memory-centric tiling, \name not only supports the next 1000x growth in model size, but also makes large models accessible to data scientists with limited GPU resources.  

\textbf{Excellent Training Efficiency} \name introduces a novel data partitioning strategy for leveraging aggregate memory bandwidth across all devices, which we refer to as \textit{bandwidth-centric partitioning}, and combines it with powerful communication \textit{overlap-centric design}, as well as optimizations for high performance NVMe access in the infinity offload engine. Together, \name offers excellent training efficiency, despite offloading data to CPU or NVMe, unencumbered by their limited bandwidth. 

\textbf{Ease of Use} With \name, data scientists no longer have to adapt their model to multiple forms of parallelism like in 3D parallelism. This is possible due to \textit{memory-centric tiling} in \name discussed above aimed at reducing GPU memory requirements of large individual layers that would otherwise require model parallelism (tensor-slicing) to fit the layers in GPU memory. In addition, \name eliminates the need for manual model code refactoring, even when scaling to trillions of parameters via an \textit{ease inspired implementation} that automates all of the communication and data partitioning required for training arbitrary model architectures. 

The main contributions of this paper are as follows:
\begin{itemize}[leftmargin=*]
\item Memory and performance characterization for large model training that describes the memory requirements (\cref{sec:memory-requirements}) for different components of a large model training as well as their bandwidth requirements (\cref{sec:bandwidth-requirements}) for the training to be efficient. 
\item \name (\cref{sec:overview}, \ref{sec:deepdive} \& \cref{sec:ease-inspired-implementation}): A novel DL training system technology consisting five innovative technologies to address the memory and bandwidth requirements for offering unprecedented model scale that is accessible and easy to use while achieving excellent training efficiency: i) \textbf{infinity offload engine} to fully leverage heterogeneous architecture on modern clusters by simultaneously exploiting GPU, CPU and NVMe memory, and GPU and CPU compute, ii) \textbf{memory-centric tiling} to handle massive operators without requiring model parallelism, iii) \textbf{bandwidth-centric partitioning} for leveraging aggregate memory bandwidth across all parallel devices, iv) \textbf{overlap-centric design} for overlapping compute and communication, v) \textbf{ease-inspired implementation} to avoid model code refactoring.
\item An extensive evaluation of \name demonstrating: i) unprecedented scale running 32 trillion parameters on 32 NVIDIA DGX-2 nodes (512 V100 GPUs), ii) excellent training efficiency achieving over 25~petaflops in throughput on the same hardware, iii) superlinear scalability of a trillion parameter model, iv) accessibility and ease-of-use: fine-tune up to a trillion parameter model on a single DGX-2 node, without using any model parallelism or model code refactoring, and v) impact of different technologies in \name on model-scale and efficiency (\cref{sec:evaluation}).
\item A discussion of \name and its potential implications of future hardware system design (\cref{sec:implication})
\item An open source implementation of \name in \href{https://www.deepspeed.ai/}{DeepSpeed}\footnote{\url{https://www.deepspeed.ai/}}, a deep learning optimization library for making distributed training easy, efficient, and effective training that has been extensively adopted by the DL community.

\end{itemize}

\section{Background and Related Work}
\label{sec:related-work}

\textbf{Data, Model, Pipeline and 3D Parallelism} Parallelization is an important strategy for training large models at scale. For a model that fits in the device memory for training, data parallelism (DP) can be used to scale training to multiple devices. 
When models do not fit in device memory, model parallelism\footnote{In this paper, we make a distinction between model parallelism and pipeline parallelism, where the former is limited specifically to mean tensor-slicing based approaches, and does not include pipeline parallelism.} (MP) \cite{DBLP:journals/corr/mesh-tensor, megatronlm, 10.1145/3302424.3303953} and pipeline parallelism (PP) \cite{GPipe, DBLP:journals/corr/pipedream, megatronlm} can split the model among processes, vertically and horizontally, respectively. 
3D parallelism \cite{deepspeed3dblog,megatrongithub} combines data, model, and pipeline parallelism to leverage the merits of each, allowing it to scale to trillions of parameters efficiently. While 3D parallelism can be highly efficient, it requires i) significant model code refactoring to split the model into model and pipeline parallel components, ii) models with complex dependency graphs are difficult to be expressed into load-balanced pipeline stages and iii) the model size is limited by the total available GPU memory.
We refer the reader to Ben-Nun and Hoefler~\cite{ben2019demystifying} for a thorough survey on parallelism in DL.

\textbf{ZeRO: Zero Redundancy Optimizer}
\label{sec:related-work-zero3}
ZeRO~\cite{zero} removes the memory redundancies across data-parallel processes by partitioning the three model states (i.e., optimizer states, gradients, and parameters) across data-parallel processes instead of replicating them. By doing so, it boosts memory efficiency compared to classic data parallelism while retaining its computational granularity and communication efficiency. There are three stages in ZeRO corresponding to three model states: the first stage (ZeRO-1) partitions only the optimizer states, the second stage (ZeRO-2) partitions both the optimizer states and the gradients, and the final stage (ZeRO-3) partitions all three model states. 
In ZeRO-3, the parameters in each layer of the model are owned by a unique data parallel process. During the training, ZeRO-3 ensures that the parameters required for the forward or backward pass of an operator are available right before its execution by issuing broadcast communication collectives from the owner process. After the execution of the operator, ZeRO-3 also removes the parameters as they are no longer needed until the next forward or backward pass of the operator. Additionally, during the parameter update phase of training, ZeRO-3 ensures that each data-parallel process only updates the optimizer states corresponding to the parameters that it owns. Thus, ZeRO-3 can keep all the model states partitioned throughout the training except for the parameters that are required by the immediate computation.  

\textbf{Heterogeneous Training Approaches} Out of several heterogeneous CPU memory based training approaches~\cite{hildebrand2020autotm,huang2020swapadvisor,jin2018layer,peng2020capuchin,ren2021sentinel,rhu2016vdnn,wang2018superneurons}, \zoff~\cite{zero-offload} is the state-of-the-art (SOTA) for large model training on multi-GPUs. \zoff is built on top of ZeRO-2 and stores the gradients and the optimizer states in CPU memory. 
\zoff leverages CPU memory in the absence of enough GPU devices to store the optimizer states and gradients. However, it still requires the parameters to be stored in GPU memory and replicated across all devices. Thus, the model scale with \zoff is limited to the total number of parameters that the memory on a single GPU device can host. \zoff also requires a large batch size to remain efficient due to suboptimal data partitioning and limited PCIe bandwidth. We address these limitations of \zoff with \name. 
In terms of NVMe based approaches, Zhao~et~al.~\cite{hierarchical-parameter-server} use a hierarchical parameter server-based design to offload sparse parameters to SSD for creating a massive scale DL Ads System. In contrast, \name is designed to be a generic DL system for training massive dense models.

\textbf{Reducing Activation Memory} Activations are the intermediate results produced during the forward propagation that need to be retained to compute the gradients during backward propagation. Multiple efforts have focused on reducing the memory required by activations through compression~\cite{jain2018gist}, activation checkpointing~\cite{DBLP:journals/corr/ChenXZG16, Jain2019CheckmateBT}, or live analysis \cite{DBLP:journals/corr/abs-1801-04380}. \name works together with activation checkpointing to reduce activation memory.  

\textbf{Adam Optimizer and Mixed Precision Training } Adaptive optimization methods~\cite{10.5555/Adagrad,DBLP:journals/corr/Adam,DBLP:journals/corr/You-LARS,DBLP:journals/corr/You-LAMB} are crucial to achieving SOTA performance and accuracy for effective model training of large models.  Compared to SGD, by maintaining fine-grained first-order and second-order statistics for each model parameter and gradient at the cost of significant memory footprint. Adam \cite{DBLP:journals/corr/Adam} is the optimizer used most prominently in large model training. 

Large model training is generally trained in mixed precision, where the forward and backward propagation are done in FP16 and the parameter updates in FP32~\cite{micikevicius2017mixed}. This leverages the performance acceleration of the tensor core units available on modern GPUs~\cite{tensor-cores}.
\vspace{-5mm}
\section{Memory Requirements}
\label{sec:memory-requirements}

\begin{figure*}
     \centering
     \begin{subfigure}[b]{0.45\textwidth}
         \centering
         \includegraphics[width=\textwidth]{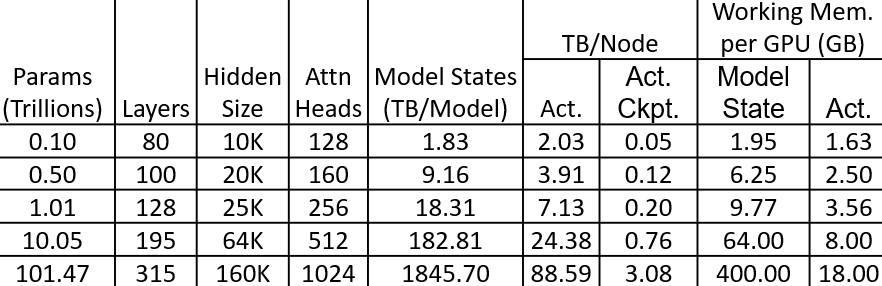}
         \caption{}
         \label{fig:model_mem_rqmts}
     \end{subfigure}
     \hfill
     \begin{subfigure}[b]{0.45\textwidth}
         \centering
         \includegraphics[width=\textwidth]{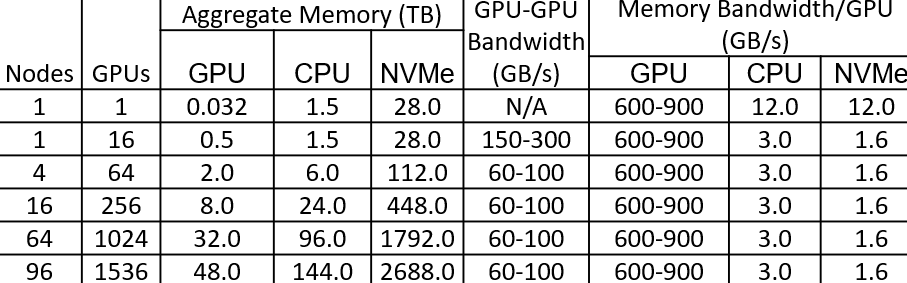}
         \caption{}
         \label{fig:DGX2_memory_and_bandwidth}
     \end{subfigure}
     \hfill
        \vspace{-1.5em}
        \caption{(a) Memory requirements for massive models. (b) Available memory and achievable bandwidth on NVIDIA V100 DGX-2 Cluster (The reported bandwidths represent per GPU bandwidth when all GPUs are reading data in parallel from the designated memory).}
        \label{fig:merged_mem_profile}
        \vspace{-2mm}
\end{figure*}

This section characterizes the memory requirements for DL training. While our methodology is generic, we focus the concrete analysis on Transformer~\cite{DBLP:journals/corr/attention} based architectures since all of the SOTA models with over a billion parameters follow that.  Our analysis assumes mixed precision training with the Adam optimizer since this recipe is the de facto standard for training Transformer based models.

The memory required for training can be categorized into two components: i) Model states including optimizer states, gradients, and model parameters, ii) Residual states primarily referring to activation memory.  To study training on heterogeneous resources, we also characterize the GPU working memory, describing the minimum amount of memory that must be available on the GPU to support training, assuming the model and residual states can be successfully offloaded from GPU memory.

\textbf{Memory for Model States:} 
The model states are comprised of optimizer states, gradients, and parameters. For mixed precision training with Adam optimizer, the parameters and gradients are stored in FP16 while the optimizer states consist of FP32 momentum, variance, parameters, and gradients. In total, each parameter requires 20 bytes of memory. The total number of parameters in a Transformer based model primarily depends on the hidden dimension ($hd$) and the number of Transformer layers ($nl$). Nearly all the parameters in a Transformer block come from four linear layers within each block with sizes: $(hd,3hd)$, $(hd, hd)$, $(hd, 4hd)$ and $(4hd, hd)$, respectively. Thus, the total parameters in a Transformer based model and can be approximated as 
\begin{equation}
12 \times nl \times hd^2
\end{equation}
requiring a total memory
\begin{equation}
240 \times nl \times hd^2
\end{equation}
in bytes to store the model states.

\Cref{fig:model_mem_rqmts} column 5 shows the memory required to store the model states of a GPT-3 like Transformer based model with 100 billion to a 100 trillion parameters created by varying hidden dimension and number of layers. To put the memory requirements in context, \Cref{fig:DGX2_memory_and_bandwidth} column 3 shows the aggregate GPU memory available on a single NVIDIA V100 DGX-2 box as well as a DGX-2 SuperPOD cluster. Note that it requires 64 GPUs to just fit the model states for a 100B parameter model. Fitting a trillion parameter model requires over 512 GPUs, while a 10 trillion parameter model is beyond the scope of even a massive 1536 GPU cluster.

\textbf{Memory for Residual States: } 
The residual states primarily consist of the activation memory, which depends on the model architecture, batch size ($bsz$) and sequence length ($seq$), and it can be quite large.
On the positive side, the memory required for activation can be significantly reduced via activation checkpointing~\cite{DBLP:journals/corr/ChenXZG16}, which trades off activation memory at the expense of $0.33$x additional recomputation. Large models such as Turing-NLG 17.2B and GPT-3 175B were all trained using activation checkpointing.  The memory required to store activation checkpoints is estimated as 
\begin{equation}
\label{eq:activation_checkpoints}
2 \times bsz \times seq \times hd \times nl / ci
\end{equation}
bytes where $ci$ is the number of Transformer blocks between two activation checkpoints, and $bsz\times seq \times hd$ is the size of the input to each Transformer block.  
\Cref{fig:model_mem_rqmts} column 7 shows the memory required to store activation checkpoints for batch size of 32 and sequence length of 1024 assuming we store one activation per Transformer block. Many modern GPU clusters have 8-16 GPUs per node, and so we chose a batch size of 2-4 per GPU, resulting in a batch size of 32 as a conservative estimate of activation within each node. While the resulting activation checkpoints are orders of magnitude smaller than the full set of activations (column 6) , beyond a trillion parameters they still get too large to fit in GPU memory for the batch size and sequence length under consideration.

\textbf{Model State Working Memory (MSWM)} is the minimum amount of GPU memory required to perform forward or backward propagation on the largest single operator in the model after all the model states have been offload to CPU or NVMe. This is approximately given by the size of the parameters and gradients of that operator in the model, since there must be at least enough memory to hold the parameter and its gradient for backward propagation. 

For a Transformer based model, the largest operator is a linear layer that transforms hidden states from $hd$ to $4hd$. The size of the parameter and gradients of this linear layer in bytes is 
\begin{equation}
4 \times hd \times 4hd
\end{equation}
Note that MSWM (\Cref{fig:model_mem_rqmts} Column 8) grows significantly beyond a 100 billion parameters, requiring multiple gigabytes in contiguous memory, which can result in running out of memory during training due to lack of enough contiguous memory to satisfy these requirements. State-of-art approaches like 3D Parallelism, addresses this issue via model parallelism, by splitting individual operator across multiple GPUs. In \cref{subsec:memory-centric-tiling}, we discuss a novel approach for addressing these massive model state working memory without requiring model parallelism.

\textbf{Activation Working Memory (AWM)} is the memory required in the backward propagation for recomputing the activations before performing the actual backward propagation. This is the size of the activations between two consecutive activation checkpoints. For example, if we create one activation checkpoint per Transformer block, the memory is given by the size of the total activation per Transformer block. This is given in bytes by approximately 
\begin{equation}
bsz \times seq \times ci \times ( 16 \times hd + 2 \times attn\_{heads} \times seq).
\end{equation}
\Cref{fig:model_mem_rqmts} column 8 shows that AWM gets large beyond 10 trillion parameters, even with $ci=1$. Unlike MSWM that is only composed of a single parameter and gradient, AWM is composed of dozens of activations, and does not cause memory issues due to lack of contiguous memory as long as the total AWM can fit in GPU memory.
\section{Bandwidth Requirements}
\label{sec:bandwidth-requirements}
\begin{figure*}[t!]
     \centering
     \begin{subfigure}[b]{0.32\textwidth}
         \centering
         \includegraphics[width=\textwidth]{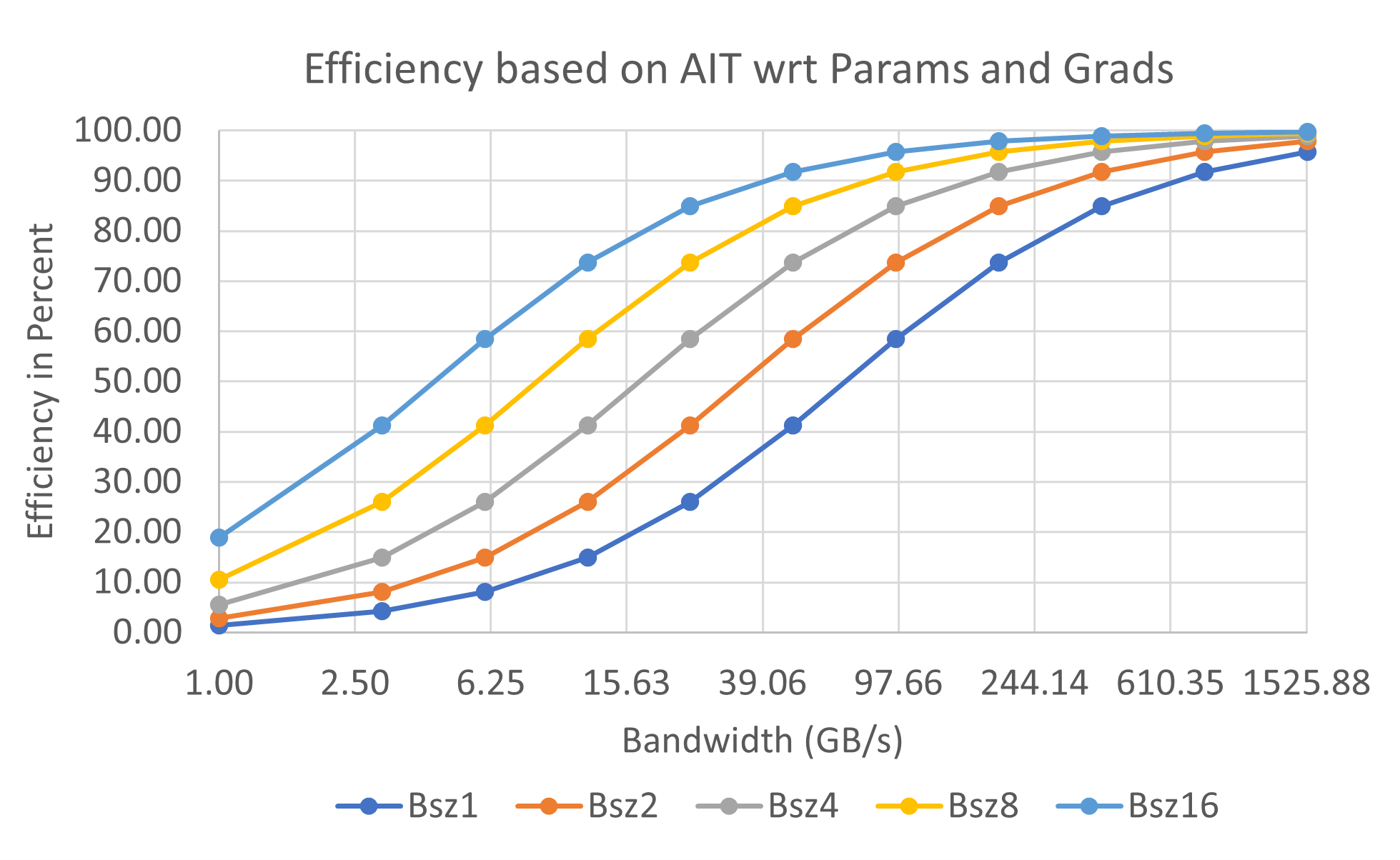}
         \caption{Parameter and Gradient Bandwidth}
         \label{fig:efficiency-wrt-param}
     \end{subfigure}
     \hfill
     \begin{subfigure}[b]{0.32\textwidth}
         \centering
         \includegraphics[width=\textwidth]{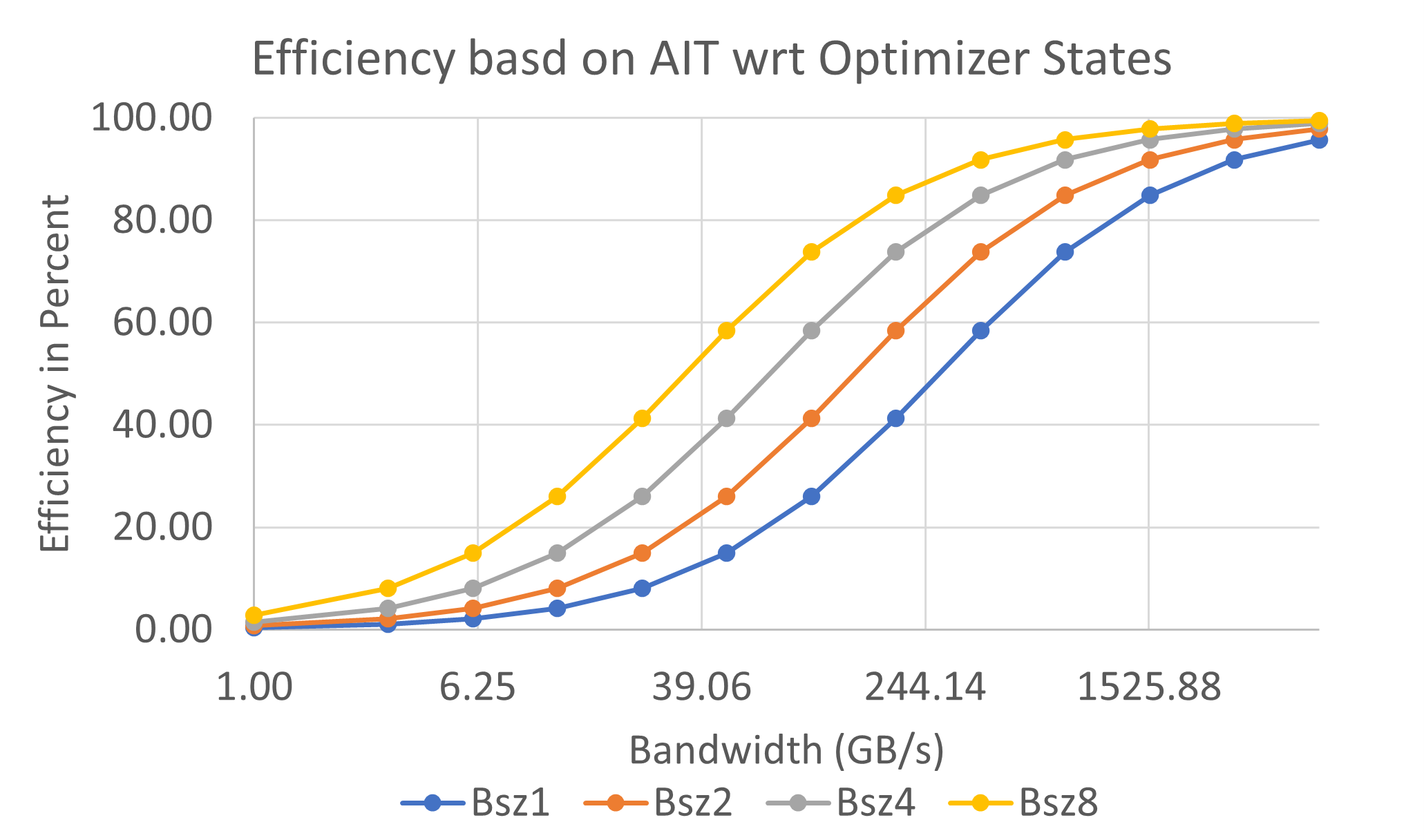}
         \caption{Optimizer States bandwidth}
         \label{fig:efficiency-wrt-os}
     \end{subfigure}
     \hfill
     \begin{subfigure}[b]{0.32\textwidth}
         \centering
         \includegraphics[width=\textwidth]{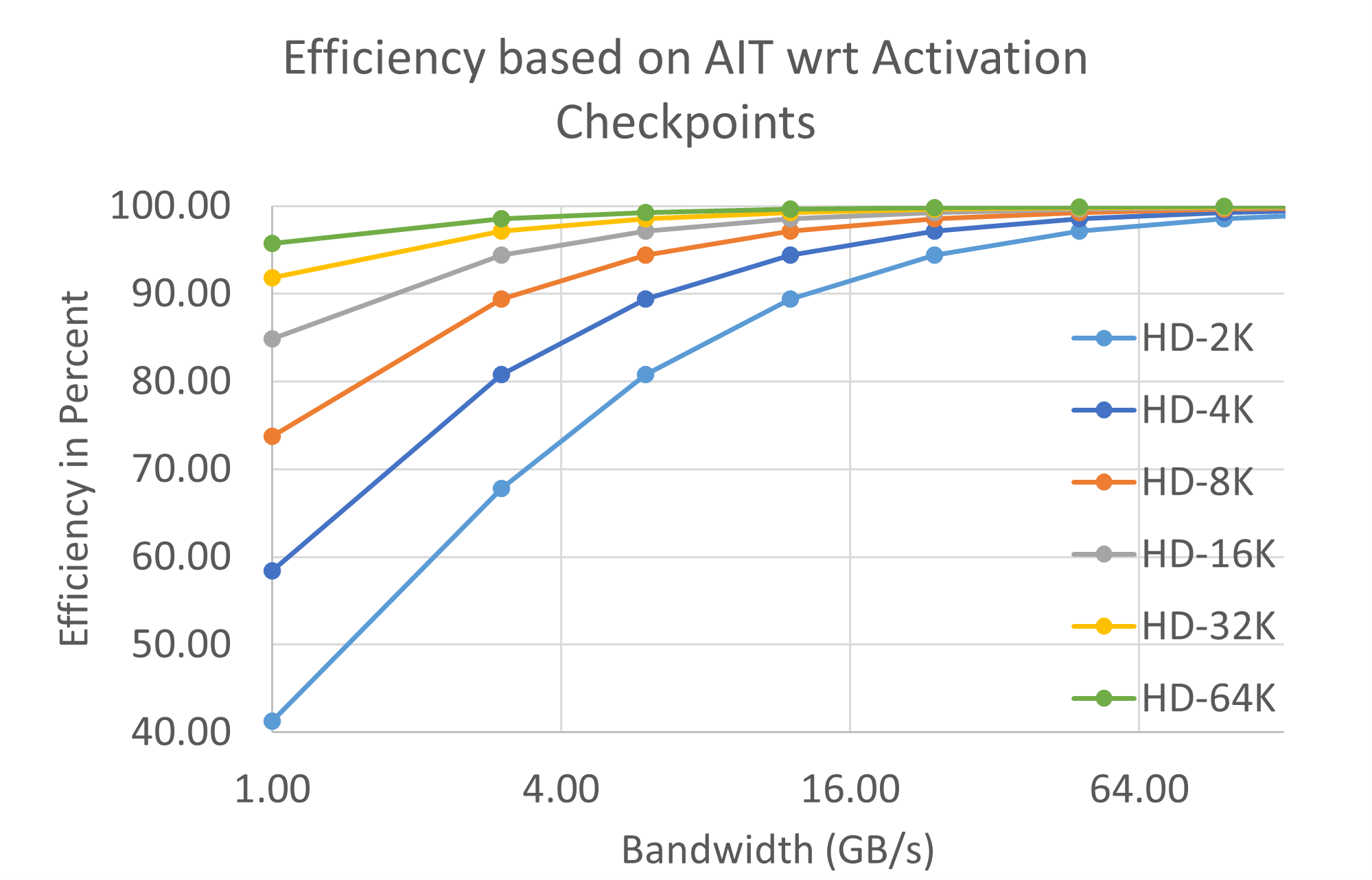}
         \caption{Activation Checkpoint Bandwidth}
         \label{fig:efficiency-wrt-activations}
     \end{subfigure}
        \vspace{-2mm}
        \caption{Impact of bandwidth on efficiency assuming an accelerator with 70 TFlops of single GPU peak achievable throughput. }
        \label{fig:efficiency-vs-bandwidth}
        \vspace{-2mm}
\end{figure*}
A critical question of offloading to CPU and NVMe memory is whether their limited bandwidth will hurt training efficiency.  This section characterizes the impact of bandwidth on training efficiency.

We start from defining an efficiency metric.  Assuming a workload execution without any compute and communication overlap, we can use the peak computational throughput ($peak_{tp}$), data movement bandwidth ($bw$) and its arithmetic intensity ($ait$) to estimate the training efficiency. 

The arithmetic intensity (AIT) of a workload is the ratio between the total computation and the data required by the computation. It describes the amount of computation per data movement. Higher AIT means a lower requirement on the data movement bandwidth, since for each data loaded the accelerator can do more computations. The efficiency metric can derived as follows:
\begin{eqnarray}
    \nonumber
    compute\_time &= \frac{total\_computation}{peak_{tp}} \\
    \nonumber
\end{eqnarray}
\begin{eqnarray}
    \nonumber
    ait &= \frac{total\_computation}{total\_data\_movement} \\
    \nonumber
    communication\_time &= \frac{total\_data\_movement}{bw} \\
    \nonumber
    &= \frac{total\_computation}{ait \times bw} \\
    \nonumber
    efficiency &= \frac{compute\_time}{compute\_time + communication\_time}
    \nonumber
\end{eqnarray}
The efficiency can be written as a function of $peak_{tp}$, $bw$ and $ait$:
\begin{equation}
    efficiency = \frac{ait \times bw}{ait \times bw + peak_{tp}}
    \label{eqn:efficiency}
\end{equation}

We will use this simple efficiency equation to characterize the data movement bandwidth required for training massive models. But before that, we will first quantify $ait$ for DL training workloads.

\subsection{Quantifying AIT in DL training}
\label{sec:quantify_ait}
Model states and activation checkpoints can have varying $ait$. We can quantify them by first identifying the total computation in each iteration of DL training, and then identifying the data movement volume for each of the model states and activations.

\noindent\textbf{Total Computation per Iteration} The total computation per iteration is dominated by the computation in the linear layers of the Transformer. For the forward propagation this can be approximated as a function of the number of parameters, sequence length, and batch size, given by $2 \times bsz \times seq \times params$. The cost of backward propagation is approximately twice that of forward propagation. Additionally, activation checkpointing requires an additional forward computation as part of recomputation during backward propagation. Therefore, the total computation per iteration is:
\begin{eqnarray}
\label{eqn:total_comp1}
computation\_per\_iter &= 2 \times 4 \times bsz \times seq \times parameters \\                              \label{eq:total_comp2} 
                   &= 2 \times 4 \times 12 \times bsz \times seq \times nl \times hd^2
\end{eqnarray}

\noindent\textbf{AIT w.r.t. Parameters and Gradients} During forward and back propagation, model \textit{parameters} must be loaded from the source location to GPU registers at least twice, i) during forward, ii) during the actual backward, resulting in a data movement of $2 \times parameters$. In presence of activation checkpointing, the parameters may be loaded one additional time for re-computation during the backward pass, adding another $1 \times parameters$. Furthermore, the \textit{gradients} must be stored from the GPU registers to its final location at least once, adding a final $1 \times parameters$ in data movement.

Therefore, assuming that parameters and gradients are stored at the same final location, the total data movement during the forward and backward pass would be $4 \times parameters$, i.e. $2 \times 4 \times parameters$ in bytes. The total computation per iteration is given by \cref{eqn:total_comp1}.  Therefore the $ait$ w.r.t \textit{parameter and gradients} is 
\begin{equation}
seq \times bsz.
\end{equation}
\noindent\textbf{AIT w.r.t. Optimizer States} During the optimizer step, the optimizer states must be read at least once, and the optimizer states must be written at least once. So the total data movement is $2 \times optimizer\_states$, which is approximately $2 \times 16 \times parameters$ bytes. The total computation per iteration is given by \cref{eqn:total_comp1}. Therefore $ait$ w.r.t \textit{optimizer states} during a full training iteration is 
\begin{equation}
seq \times bsz / 4.
\end{equation}
\noindent\textbf{AIT w.r.t. Activation Checkpoints}
During the forward propagation activation checkpoints must be saved to their final location, and must be retrieved during the backward propagation. Therefore, the total data movement w.r.t activation checkpoints in bytes is given by $2 \times total\_activation\_checkpoints\_in\_bytes$ which is given by $4 \times nl/ci \times hd \times seq \times bsz$ from \cref{eq:activation_checkpoints}. The total computation per iteration is given by \cref{eq:total_comp2}. So the $ait$ w.r.t activation checkpoints is given by 
\begin{equation}
24 \times hd \times ci.
\end{equation}
\subsection{Bandwidth Requirements}
Due to the variation in the AIT, model states and activation checkpoints have very different bandwidth requirements to achieve good efficiency. The former only depends on the batch size and sequence length, while the latter only depends on the frequency of activation checkpoints and hidden dimension size of the model. 

Besides AIT, the bandwidth requirement for efficiency also depends  on $peak_{tp}$, as shown in \cref{eqn:efficiency}. Using $peak_{tp}$, and $ait$ we first show how efficiency varies with bandwidth w.r.t to different model and residual states, and then discuss the bandwidth requirements on these states for DL training to be efficient. Our methodology is generic and can be applied to understanding the bandwidth requirements on any current or future generation clusters. Here, we use NVIDIA V100 DGX-2 SuperPOD cluster as our example platform.

Using the $ait$ expression from \cref{sec:quantify_ait} and efficiency metric based on \cref{eqn:efficiency}, \Cref{fig:efficiency-vs-bandwidth} shows the relationship between efficiency and available bandwidth w.r.t. parameter and gradients, optimizer states, and  activation checkpoints. To produce these plots, we computed the $ait$ based on expressions derived in \cref{sec:quantify_ait}, for varying batch sizes, sequence length and model configurations. More specifically, we use a sequence length of 1024, the same sequence length used for GPT-2~\cite{gpt-2}, Megatron-LM~\cite{megatronlm}, and Turing-NLG~\cite{turing-nlg17B}. We vary batch size range from 1 to 16 to capture large GPU and small GPU experiments, respectively. A small batch size per GPU is used when running on large number of GPUs, while a large batch size per GPU is used when training on relatively fewer GPUs to maintain a reasonable effective batch size for training. Our hidden size ranges from 8K-64K representing models with hundreds of billions of parameters, to tens of trillions of parameters as shown in \Cref{fig:model_mem_rqmts}. 

To identify $peak_{tp}$ for this analysis, we use an empirical approach\footnote{Note that $peak_{tp}$ is not the theoretical hardware peak, but instead the achievable peak in the absence of any communication bottleneck.}. We ran models with aforementioned configurations on a single NVIDIA V100 DGX-2 box with all non-GPU communication turned off to simulate a virtually unlimited bandwidth scenario. The performance achieved ranged from 62-78 TFlops/GPU based on the hidden size of 8K-64K, respectively. We used the average of 70 TFlops/GPU to represent $peak_{tp}$ for the purpose of this analysis\footnote{Results will vary based on the value of $peak_{tp}$ used, and this analysis is a single data point, meant as a guide for understanding relationship between efficiency and bandwidth for DL workloads specifically on the NVIDIA V100 DGX-2 clusters. Furthermore, the result only considers the relationship between efficiency and bandwidth of model states and activations, one at a time, assuming infinite bandwidth for others to isolate the bandwidth requirement for each state separately.}.

\textbf{Bandwidth w.r.t. Parameter and Gradients} \Cref{fig:efficiency-wrt-param} shows that with a bandwidth of over 70 GB/s for parameter and gradients, we can achieve over $50\%$ efficiency for even the smallest batch size. At this bandwidth, the data movement in theory can be completely overlapped with the computation to achieve a $100\%$ efficiency. 

\textbf{Bandwidth w.r.t. Optimizer States}
\Cref{fig:efficiency-wrt-os} shows that optimizer states require nearly 4x higher bandwidth to achieve $50\%$ efficiency compared to parameters and gradients. Furthermore, the optimizer states are updated at the end of the forward and backward propagation and cannot be overlapped with the computation. As a result they require significantly larger bandwidth to keep the overall DL workload efficient. For example achieving $90\%$ efficiency with batch size of 2 per GPU requires nearly 1.5 TB/s of effective bandwidth, which is greater than even the GPU memory bandwidth.

\textbf{Bandwidth w.r.t. activation memory}
\Cref{fig:efficiency-wrt-activations} also shows that with activation checkpointing enabled, a meager bandwidth of 2~GB/s is able to sustain over $50\%$ efficiency even for a hidden size of $2K$. The bandwidth requirement drops down to less than 1~GB/s once the hidden size grows over $8K$.

\section{\name Design Overview}
\label{sec:overview}
\begin{figure}[t]
    \centering
    \includegraphics[width=\columnwidth]{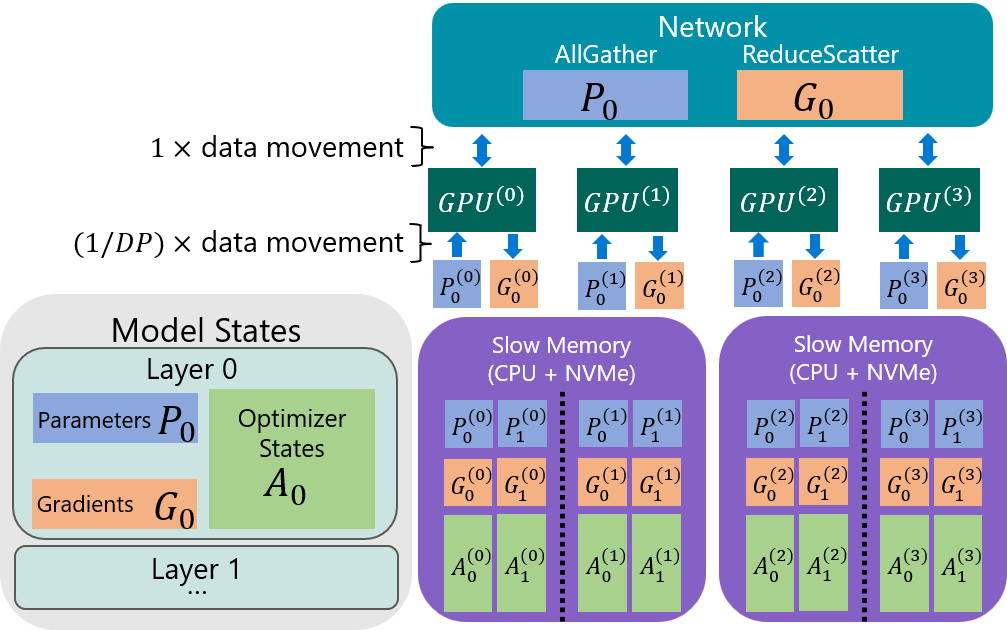}
    \caption{
        \small
        A snapshot of \name training a model with two layers on four data parallel (DP) ranks. Communication for the backward pass of the first layer is depicted. Partitioned parameters are moved from slow memory to GPU and then collected to form the full layer. After gradients are computed, they are aggregated, re-partitoned, and then offloaded to slow memory. Layers are denoted with subscripts and DP ranks are denoted with superscripts. For example, $P_{0}^{(2)}$ is the portion of layer 0's parameters owned by $GPU^{(2)}$.
    }
    \label{fig:zero-infinity-overview}
\end{figure}

In this section we present an overview of the design choices in \name that enable it to achieve unprecedented model scale while offering excellent training efficiency and ease of use. A bird's eye view of \name is illustrated in \Cref{fig:zero-infinity-overview} and discussed below.

\subsection{Design for Unprecedented Scale} 
Modern GPU clusters are highly heterogeneous in terms of memory storage. In addition to the GPU memory, they have CPU memory as well as massive NVMe storage that is over 50x larger than the GPU memory and nearly 20x larger than CPU memory (See \cref{fig:DGX2_memory_and_bandwidth}). 

We developed \name, a parallel system for DL training that can transcend the GPU memory wall by exploiting these heterogeneous memory systems in modern GPU clusters. \Cref{fig:max-model-size-128nodes} compares the maximum achieved model size of 3D parallelism and \name. \name supports one trillion parameters per NVIDIA~V100~DGX-2 node, a 50x increase over 3D~parallelism.

\subsubsection{Infinity offload engine for model states} \name is built on top of ZeRO-3~\cite{zero} which partitions all model states to remove memory redundancy as discussed in \cref{sec:related-work-zero3}. Unlike any of the existing ZeRO family of technology, \name is designed with a powerful offload mechanism called the infinity offload engine which can offload all of the partitioned model states to CPU or NVMe memory, or keep them on the GPU based on the memory requirements. Note from \cref{fig:model_mem_rqmts} and \cref{fig:DGX2_memory_and_bandwidth}, even the model states required by a 100 trillion parameter model can fit in the aggregate NVMe memory of a DGX-2 cluster with 96 nodes (1536 GPUs). Therefore, the infinity offload engine allows \name to fit model states of models with hundreds of trillions of parameters. See \cref{sec:deepdive} for more details.

\subsubsection{CPU Offload for activations} In addition to model states, \name can offload activation memory to CPU memory, when necessary. Note that the activation checkpoints (0.76 TB) required by a 10 trillion parameter model can easily fit in the 1.5TB of CPU memory available on a DGX-2 system, while the 3~TBs of activation checkpoints required by a 100 trillion parameter is within reach of the CPU memory of the next generation hardware. Therefore, by offloading activation checkpoints to CPU memory, \name can fit the activation checkpoints of models with hundreds of trillions of parameters.

\subsubsection{Memory-centric tiling for working memory}
\label{subsec:memory-centric-tiling}
To reduce the working memory requirements of DL training for large models, \name introduces a novel technique called {\it memory-centric tiling} that exploits the data fetch and release pattern of ZeRO-3 to reduce the working memory requirements by breaking down a large operator into smaller tiles that can be executed sequentially.

For example, to reduce the working memory for a large linear operator, \name represents the operator as a mathematically equivalent sequence of smaller linear operators consisting of tiles of parameters from the original operator, and executes them sequentially. When combined with ZeRO-3, the parameter and gradients of each tile can be fetched and released one at a time, reducing the working memory proportional to the number of tiles. Therefore, \name can support operators of arbitrary sizes, without relying on model parallelism to fit them in limited GPU memory.

\subsection{Design for Excellent Training Efficiency}

Offloading all model states and activations to CPU or NVMe is only practical if \name can achieve high efficiency despite the offload. In reality this is extremely challenging since CPU memory is an order of magnitude slower than GPU memory bandwidth, while the NVMe bandwidth is yet another order of magnitude slower than the CPU memory bandwidth. Furthermore, reading and writing to these memory from GPU is even slower (see \cref{fig:DGX2_memory_and_bandwidth}). 

On a system like the DGX-2, the bandwidth must be greater than 70GB/s, 1.5TB/s, and 1-4 GB/s w.r.t. parameter and gradients, optimizer states, and activation checkpoints, respectively for DL training to be efficient, based on our analysis in \cref{sec:bandwidth-requirements}. Here we discuss how \name achieves the necessary bandwidths to achieve excellent efficiency.

\subsubsection{Efficiency w.r.t Parameter and Gradients}
The data movement bandwidth for parameters and gradients must be greater than 70GB/s, close to the GPU-GPU bandwidth available on DGX-2 clusters~\cite{dgx-2-infiniband-bandwidth}. Therefore, a  DL parallel training solution like ZeRO-3~\cite{zero} where parameters are broadcasted from the owner GPU to the rest before using them in forward or backward propagation can run efficiently as long as the communication is overlapped. 

On the contrary, a meager 12 GB/s PCIe bandwidth from a single GPU to CPU memory or NVMe (see \cref{fig:DGX2_memory_and_bandwidth}) or vice-versa is simply not sufficient to support heterogeneous training at scale\footnote{CPU and NVMe bandwidth are in the order of 100 GB/s and 25 GB/s, respectively, but reading data from CPU or NVMe to a single GPU is limited by the achievable PCIe bandwidth which is around 10-12 GB/s}. Therefore, existing heterogeneous solutions like \zoff where the parameters must be first moved from CPU to owner GPU before broadcasting requires significantly large batch sizes per GPU to achieve enough $ait$ necessary to be efficient under the limited bandwidth. This poses two problems: i) for massive models the activation memory will get too large to fit even in CPU memory, and ii) the effective batch size becomes too large when scaling to hundreds or thousands of GPUs for effective convergence.

\name addresses these challenges in two ways: 
i) \textit {bandwidth-centric partitioning}: a novel  data  mapping  and  parallel data retrieval strategy for offloaded parameters and gradients that allows \name to achieve virtually unlimited heterogeneous memory bandwidth (details in  \cref{subsec:bandwidth-centric-partitioning}), and ii) an \textit{overlap centric design} that allows \name to overlap not only GPU-GPU communication with computation but also NVMe-CPU and CPU-GPU communications over the PCIe (details in \cref{subsec:memory-centric-tiling}).  

\subsubsection{Efficiency w.r.t Optimizer States}
Unlike parameters and gradients that are consumed and produced sequentially during the forward and backward propagation, optimizer states can be updated in parallel, all at once. This property is leveraged by both ZeRO-3 and \zoff, that store and update the optimizer states in GPU and CPU memory, respectively, in parallel across all available GPUs and CPUs. As a result the aggregate GPU or CPU memory bandwidth can get much higher than the required 1.5TB/s with increase in GPU or CPU count.

Since \name is built upon ZeRO-3, it can also leverage the aggregate GPU and CPU memory bandwidth as well as the aggregate CPU compute for optimizer step, when offloading optimizer states to CPU memory. However, with NVMe offload, it is necessary to bring the data from NVMe to CPU memory and back in chunks that can fit in the CPU memory to perform the optimizer step, one chunk at a time. The optimizer step is therefore limited by the NVMe-CPU memory bandwidth: while \name can achieve aggregate NVMe bandwidth across multiple nodes, it is crucial to achieve near peak NVMe bandwidth per node, to allow supporting the necessary bandwidth of over 1.5 TB/s with as few nodes, and as small batch size as possible. Furthermore, the process of bringing data in and out of NVMe to CPU memory, or from CPU memory to GPU memory can cause CPU memory fragmentation in both GPU and CPU that can result in out of memory even with plenty of memory still available.

The \textit{infinity offload engine} can not only achieve near peak NVMe bandwidth, it can also allows \name to overlap NVMe to CPU reads with CPU to NVMe writes, as well as the CPU computation for the optimizer step at the same time to allow \name to remain efficient with a modest batch size on small number of GPUs and with small batch sizes on large numbers of GPUs. At the same time, it minimizes memory fragmentation by carefully reusing temporary buffers for data movement. We discuss the optimizations in infinity offload engine and  in detail in \cref{sec:deepdive}.

\subsubsection{Efficiency w.r.t Activations}
On a DGX-2 node, each GPU can read and write data at about 3 GB/s to CPU memory in parallel over the PCIe allowing activation checkpoints to be offloaded to CPU memory while retaining over $80\%$ efficiency for hidden size larger $8K$ or larger. To also allow for high efficiency at smaller hidden sizes, \name can decrease the frequency of activation checkpoints as well as effectively overlap the communication of activation checkpoints both to and from CPU memory with the forward and backward computation on the GPU.

\subsection{Design for Ease of Use}
\label{sec:ease-of-use}
With \name, data scientists no longer have to adapt their model to multiple forms of parallelism like in 3D parallelism. This is possible due to \textit{memory-centric tiling} in \name discussed in \cref{subsec:memory-centric-tiling} aimed at reducing GPU memory requirements of large individual layers that would otherwise require model parallelism (tensor-slicing) to fit the layers in GPU memory. 

In addition, \name is implemented in PyTorch in a way that eliminates the need for manual model code refactoring even when scaling to trillions of parameters. This is made possible through an \textit{ease-inspired implementation} with two automated features:

i) \textit{automated data movement} to gather and partition parameters right before and after they are required during the training. \name does this by injecting i) pre forward/backward hooks into PyTorch submodules that trigger allgather collectives to collect the parameters required before its forward/backward pass and ii) post forward/backward hooks that trigger parameter/gradient partiitoning and optionally offloading them to CPU or NVMe (see \cref{sec:automatic-data-movement} for details). 

ii) \textit{automated model partitioning during initialization} such that models that can not fit within single GPU or CPU memory can still be initialized without requiring manual partitioning of the model across data parallel processes. \name achieves this by wrapping the constructor of all module classes so that parameters of each submodule are partitioned and offloaded immediately after they are created during initialization. The entire model is never fully instantiated on a single data parallel process (see \cref{sec:automatic-initialization} for details). 
\section{Efficiency Optimizations}
\label{sec:deepdive}

In this section, we deep dive into the optimizations introduced in \cref{sec:overview} that allow \name to achieve excellent efficiency. 
\subsection{Bandwidth-Centric Partitioning}
\label{subsec:bandwidth-centric-partitioning}
\name implements a novel data mapping and retrieval strategy to address the NVMe and CPU memory bandwidth limitations. Unlike ZeRO \cite{zero} and \zoff~\cite{zero-offload}, where parameters of each layer are owned by a single data parallel process, which broadcasts them to the rest when needed, \name partitions individual parameters across all the data parallel process, and uses an allgather instead of a broadcast when a parameter needs to be accessed. Note that both broadcast and allgather communication collectives have the same communication cost when it comes to data movement volume if the data is located on the GPU. Therefore, this makes no difference for a GPU-only training. However, this is a game changer when the data is located in NVMe or CPU.

In the broadcast-based approach, since each parameter is fully owned by one of the data parallel processes, the parameter must be first communicated from its source location to the GPU memory via the PCIe before the broadcast can happen. Note that only a single PCIe can be active for this process, while all the PCIe links connected to all the other GPUs are idle. On the contrary, with the partitioned parameter and allgather based approach in \name, all PCIe links are active in parallel, each bringing in $1/dp^{th}$ portion of the parameter where $dp$ is the data parallel degree. As a result, the effective communication bandwidth between NVMe or CPU to the GPU, increases linearly with the $dp$ degree.

For example, with broadcast-based approach, the CPU/NVMe to GPU bandwidth stays constant at about 12 GB/s with PCIe Gen 3, even with 16-way data parallelism on the DGX-2 box. However, with the all-gather-based approach, the effective achievable bandwidth increases to about 48/25 GB/s (3.0/1.6 GB/s per GPU), respectively (see \cref{fig:DGX2_memory_and_bandwidth}), limited only by the max aggregate PCIe bandwidth and max NVMe bandwidth per DGX-2 node. From here, the bandwidth grows linearly with more nodes. 
When training a massive model at massive scale, \name can therefore offer significantly more heterogeneous memory bandwidth than necessary (virtually unlimited) for the training to remain efficient. For example, on 64 DGX-2 nodes, \name has access to over 3TB/s of CPU memory bandwidth and over 1.5TB/s of NVMe bandwidth. 
\subsection{Overlap Centric Design}
\label{subsec:overlap}
While \name can leverage sufficient heterogeneous memory bandwidth on a multi-node setup, the bandwidth can still be a bottleneck on a single GPU or single node setup. Even the GPU-GPU allgather communication has a big impact on efficiency when running with a small batch size (\cref{fig:efficiency-vs-bandwidth}). Furthermore, accessing NVMe memory requires a three step process: i) read data from NVMe to CPU memory (nc-transfer), ii) copy the data from CPU memory to GPU memory (cg-transfer), iii) execute allgather to construct the full parameter on all GPUs (gg-transfer). The sequential nature of these data movements means that if done naively, the total communication time would be the sum of each of these three data movement cost, resulting in poor efficiency even if the bandwidth for data movement at each of these stages is individually sufficient.

To address these issues, \name has an overlap engine that not only overlaps GPU-GPU communication with GPU computation, but also overlaps the NVMe to CPU, and CPU to GPU communication, all at the same time. The overlap engine has two components: i) A dynamic prefetcher for overlapping the data movement required to reconstruct parameters before they are consumed in the forward or backward pass, and ii) a communication and offload overlapping mechanism for executing the data movement required by gradients in parallel with the backward computation.

The dynamic prefetcher in \name traces the forward and backward computation on that fly, constructing an internal map of the operator sequence for each iteration. During each iteration, the prefetcher keeps track of where it is in the operator sequence and prefetches the parameter requires by the future operators. The prefetcher is aware of the three step communication process, and therefore can overlap the nc-transfer for one parameter, with cg-transfer and gg-transfer of other parameters. For instance, before executing the $i^{th}$ operator, the prefetcher can invoke nc, cg, and gg-transfer for parameters required by $i+3, i+2,$ and $i+1$ operators, respectively. Note that all of these data movement can happen in parallel with the execution of the $i^{th}$ operator. Furthermore, \name can update the operator sequence map in case of dynamic workflow, allowing for appropriate prefetching even when the forward and backward propagation changes across iterations.

Similarly, in the backward pass, \name can overlap the reduce-scatter for gradients of the parameters in $(i+1)^{th}$ operator with the computation of the $i^{th}$ operator, while simultaneous transferring the partitioned gradients from the reduce-scatter of the gradients of the $(i+2)^{th}$ operator to the CPU or NVMe. 

With this powerful overlap centric design, \name hides significant portions of data movement even when training with a small number of GPUs and small batch size per GPU.

\subsection{Infinity Offload Engine}
\label{sec:infinity-offload-engine}
The infinity offload engine is composed of two main components: 

\textbf{DeepNVMe}, a powerful C++ NVMe read/write library in the infinity offload engine that supports bulk read/write requests for asynchronous completion, and explicit synchronization requests to flush ongoing read/writes. The support for asynchrony allows \name to overlap these requests with GPU/GPU or GPU/CPU communication or computation. 

Most importantly, DeepNVMe is capable of achieving near peak sequential read and write bandwidths on the NVMe storage device. It achieves this high performance through a number of optimizations, including aggressive parallelization of I/O requests (whether from a single user thread or across multiple user threads), smart work scheduling, avoiding data copying, and memory pinning. 

\textbf{Pinned memory management layer}  To ensure high performance tensor reads (or writes) from (to) NVMe/CPU storage, the source (or destination) tensors must reside in pinned memory buffers. However, pinned memory buffers are scarce system resources, and their oversubscription by a single process can degrade overall system performance or cause system instability. This layer manages the limited supply of pinned memory by reusing a small amount (tens of GBs) for offloading the entire model states (up to tens of TBs) to CPU or NVMe. The reuse of memory buffer prevents memory fragmentation in CPU and GPU memory. This layer also provides PyTorch tensors with pinned memory data, allowing in-place computation of the tensors so that they can then be written to NVMe without any further copies to improve bandwidth.

\section{Ease Inspired Implementation}
\label{sec:ease-inspired-implementation}

\name is implemented on top of PyTorch, and is designed to be used without any model code refactoring similar to standard data-parallel training in PyTorch. This section details some of the challenges faced in implementing such a system. 

\begin{table*}[tbh!]
\resizebox{1.5\columnwidth}{!}{
\begin{tabular}{|c|c|c|c|c|c|c|c|}
\hline
 \# nodes& \# params& hidden dim & \# layers & batch/GPU&  mp & \ fp16 param & Opt State  \\ \hline 
1 & 10 B        & 4K            & 50       & 8  & 1 & GPU & GPU \\ \hline
1 & 50, 100 B   & 8K            & 62, 125  & 26, 24  & 1 & CPU & NVMe \\ \hline
1 & 0.5, 1 T    & 18K, 25K      & 124, 128 & 8, 7 & 1 & NVMe    &  NVMe   \\ \hline
32 & 0.5, 1 T   & 18K, 25K      & 124, 128 & 7, 5  & 4 & GPU & GPU \\ \hline
32 & 5, 10, 20 T& 48K, 64K, 88K & 174, 200, 205 & 3, 2, 1.25 & 4, 4, 8  & NVMe    &  NVMe   \\ \hline
\end{tabular}
}
\caption{Experiment configurations. Sizes are expressed in B for billions, T for trillions, and K for 1024.}
\label{tab:merged_configs}

\end{table*}
\begin{figure*}[t!]
     \centering
     \begin{subfigure}[b]{0.32\textwidth}
         \centering
         \includegraphics[width=\textwidth]{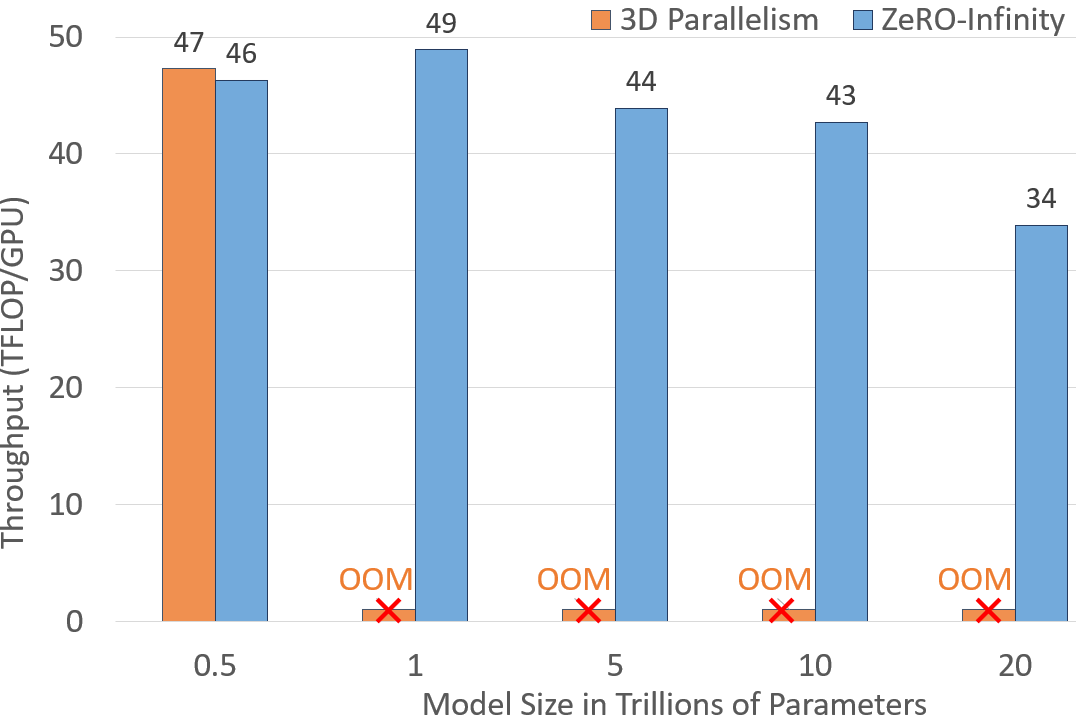}
         \caption{\name efficiently trains 40x larger models than 3D parallelism on 512 GPUs.}
         \label{fig:perf_512gpu}
     \end{subfigure}
     \hfill
     \begin{subfigure}[b]{0.32\textwidth}
         \centering
         \includegraphics[width=\textwidth]{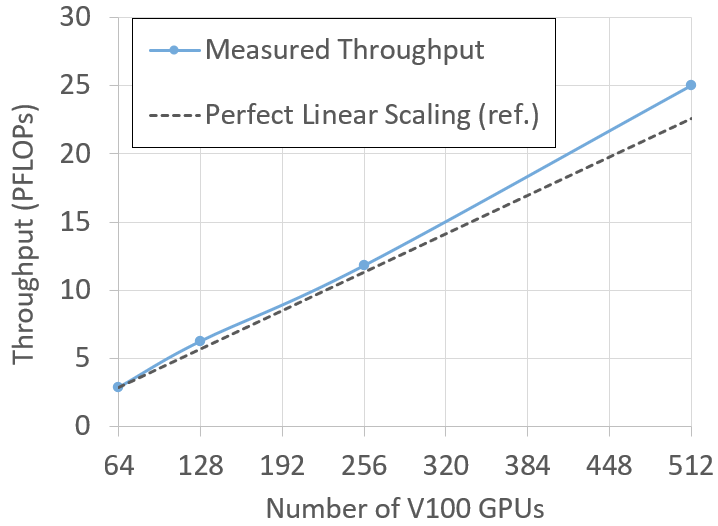}
         \caption{\name exceeds linear scaling from 64 to 512 GPUs for a 1T parameter model.}
         \label{fig:scalability_1T}
     \end{subfigure}
     \hfill
     \begin{subfigure}[b]{0.32\textwidth}
         \centering
         \includegraphics[width=\textwidth]{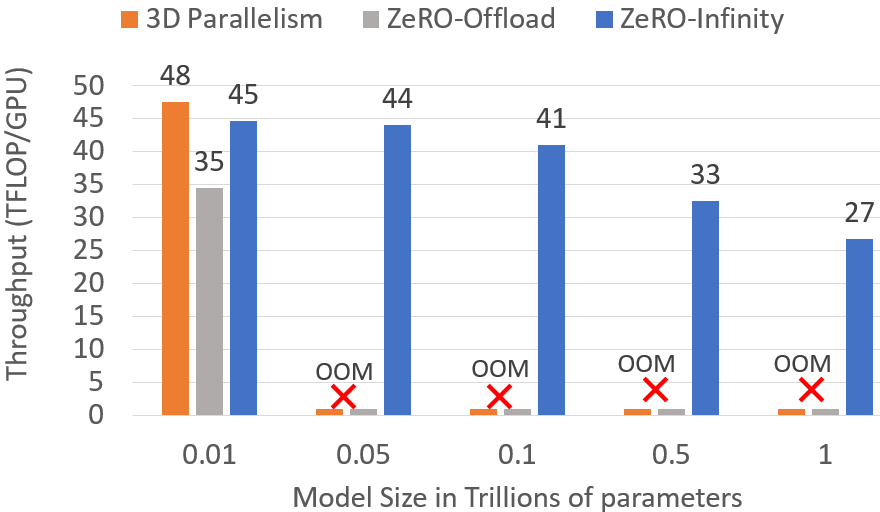}
         \caption{\name can train up to 1T model on a DGX-2 node without model parallelism.}
         \label{fig:single_node_perf}
     \end{subfigure}
        \vspace{-2mm}
        \caption{Efficiency and scalability of \name for training multi-trillion parameter models. }
        \label{fig:merged_perf}
\end{figure*}
\subsection{Automating Data Movement}
\label{sec:automatic-data-movement}
\name must coordinate the movement of tensors comprising the model parameters, gradients, and optimizer states. When a tensor is not in active use, it remains partitioned among workers and potentially offloaded to CPU or NVMe memory. The system must ensure that the tensors are resident in GPU memory in time for use and then later re-partitioned. 

PyTorch models are expressed as a hierarchy of modules that represent the layers of a neural network. For example, Transformer architectures~\cite{DBLP:journals/corr/attention} contain submodules such as self-attention and feedforward networks. The self-attention submodules are further comprised of linear transformations and other submodules. 

\name recursively injects hooks into the submodules of a model to automate the required data movement. At the start of a submodule’s forward pass, these hooks ensure that the submodule's parameters are available for computations, otherwise it will execute the appropriate allgather collectives and block until the parmaeters become available. The overlap-centric design detailed in \cref{subsec:overlap} is critical to minimizing stalls due to parameter communication. At the end of the submodule's forward pass, we partition the parameters again and optionally offload them. The backward pass is handled in a similar fashion. 

\subsubsection{Auto Registration of External Parameters}
In the ideal case, a submodule’s parameters and gradients are only accessed within its own forward and backward passes, making it straight forward to identify and automate the data movement as discussed in section above. However, some model architectures are exceptions, where parameters defined and allocated in a submodule is used in forward and backward propagation of a different submodule. For example, language models such as GPT~\cite{radford2018improving} share the weights of the embedding layer at both the beginning and the end of the network to map words to vectors and vice versa. 
We refer to the parameters that are used across module boundaries as \emph{external parameters}.  In presence of external parameters, it is difficult to know which parameters to gather at the beginning of a submodule's forward and backward pass.  

One way to address this is to \emph{register} external parameters with \name so that they are collected for the forward and backward passes of the submodule that access them. After registration, an external parameter is treated like all others and will be included in the prefetching system as described in \cref{subsec:overlap}.  We provide APIs for manual registration of external parameters.

In order to improve user experience, we also provide mechanisms to detect these scenarios and automatically register external parameters so the the user does not have to make any code change:

\textbf{Intercepting partitioned parameter accesses}
PyTorch modules store their tensor parameters in a hash table. At the initialization time, we replace the hash table with a subclassed type that overrides the tensor accesses. When a partitioned parameter is accessed, we do a blocking allgather on the parameter, register it as an external parameter, and then return the gathered parameter.

\textbf{Activation introspection} A submodule may return a parameter from its forward pass to be consumed by another submodule's forward and backward passes. For example, Megatron-LM returns bias vectors from the forward pass of linear layers and they are consumed by the parent Transformer layer modules. We inspect the activation outputs returned from each submodule's forward pass for partitioned parameters. If one is discovered, we collect and register it as an external parameter.

\subsection{Automatic Model Partitioning during Initialization}
\label{sec:automatic-initialization}
If the model is large, then it may not be possible to fully initialize the model with traditional data parallel approach, replicating it on each data parallel process before it can be partitioned for \name. For example, a 500 billion parameter model will occupy 1 TB of memory in half precision, and thus a system with 8 GPUs per node requires 8 TB of aggregate CPU or GPU memory just for the initial data parallel allocation step. This is beyond the GPU or CPU memory available on a node. 

To address this limitation, the parameters corresponding to each layer of the model must be partitioned at the time of initialization, and not after the entire model is initialized. To do this, we provide a Python \name \emph{context} which decorates the \texttt{\_\_init\_\_} method of \texttt{torch.nn.Module}, so that parameters allocated under each module/sub-module are partitioned immediately after its initialization among the group of data parallel processes. 

As a result, only individual sub-modules are fully initialized before they are partitioned, and the full model is never replicated on all the data parallel process. In the example above, the 500 billion parameter model can therefore be fully partitioned during its initialization requiring only 1 TB of aggregate CPU memory regardless of the total number of data parallel process.

\section{Evaluation}
\label{sec:evaluation}
This section evaluates ZeRO-Infinity, demonstrating that it achieves excellent training efficiency and scalability for models with tens of trillion parameters. We also show the impact of various technologies within ZeRO-Infinity on model scale and performance. 

\subsection{Methodology}

\subsubsection*{Hardware}
We conducted our experiments on a cluster of up to 512 V100 SXM3 32\,GB GPUs (32 DGX-2 nodes) with 800\,Gbps internode communication bandwidth. 

\subsubsection*{Baseline}
For experiments without model parallelism (mp), we use torch's distributed data parallel (DDP~\cite{torchddp}) as a baseline. For experiments with model parallelism, we use Megatron-LM~\cite{megatronlm}. As a baseline for each experiment we use the relevant state-of-the-art method among 3D Parallelism~\cite{3dparallelism}, ZeRO~\cite{zero}, or ZeRO-Offload~\cite{zero-offload}.

\subsubsection*{Model Configurations} 
We use GPT-like Transformer based models. We fix the sequence length to 1024 and vary the  hidden  dimension  and  number  of  layers  to  obtain  models with different number of parameters. \Cref{tab:merged_configs} provides the specific model configurations used throughout our evaluation, (see \cref{sec:appendix}) for additional configurations.

\subsection{Model Size and Speed}
\textbf{Model Size} \name trains models over 32 \emph{trillion} parameters compared to about 650B parameters with 3D parallelism, state of the art, offering a leap of 50x in model scale (\Cref{fig:max-model-size-128nodes}). 

\noindent\textbf{Model Speed}~ \Cref{fig:perf_512gpu} shows the performance of \name on up to 20 trillion parameter models on 512 GPUs. For the 500B model (close to the largest that 3D parallelism can run on these resources), \name and 3D parallelism achieve nearly identical throughput, indicating \name is on par with the training efficiency of the state-of-the-art.  When increasing the model size further, 3D parallelism simply runs out of memory, while \name trains up to 20 trillion parameter models (40x larger) with excellent throughput of up to 49 TFlops/GPU.
At the extreme-scale, \Cref{fig:perf_512gpu} shows a performance drop from 10T (43 TFlops/GPU), and 20T (34 TFlops/GPU). This drop is not due to NVMe bandwidth as both model sizes use NVMe offload, but instead due to an extremely small batch size per GPU (\Cref{tab:merged_configs}) at 20T scale as a result of limited CPU memory to store activation checkpoints. This can be improved by increasing the CPU memory or offloading activation checkpoints to NVMe in a future implementation. 
\subsection{Superlinear Scalability}
\Cref{fig:scalability_1T} shows that \name achieves super-linear scalability from 4 nodes (64 GPUs) to 32 nodes (512 GPUs) when training a 1T model.  This is a weak scaling result where we keep batch size per node as constant and increase total batch size with increased number of nodes.  \name exceeds perfect linear scaling by effectively leveraging the linear increase in aggregate PCIe and NVMe bandwidth to accelerate the offloading of parameters and optimizer states, and leveraging CPU compute from additional nodes for parameter update. In addition, \name already achieves over 2.8 petaflops (44 Tflops/GPU) with just 4 nodes, demonstrating that the aggregated NVMe bandwidth is sufficient to achieve good efficiency even at a modest scale.   


     
     

\begin{figure*}[thb!]
    \centering
    \begin{subfigure}[b]{0.19\textwidth}
        \includegraphics[width=\textwidth]{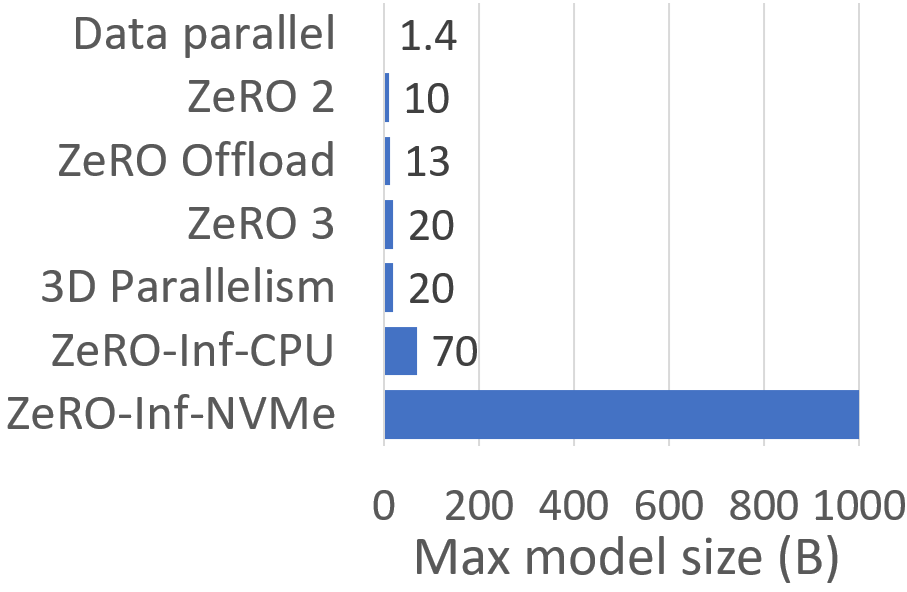}
        \captionof{figure}{\small{Max model size w.r.t. ZeRO strategies.}}
        \label{fig:max-model-size}
    \end{subfigure}
    \hfill
    \begin{subfigure}[b]{0.19\textwidth}
        \includegraphics[width=\textwidth]{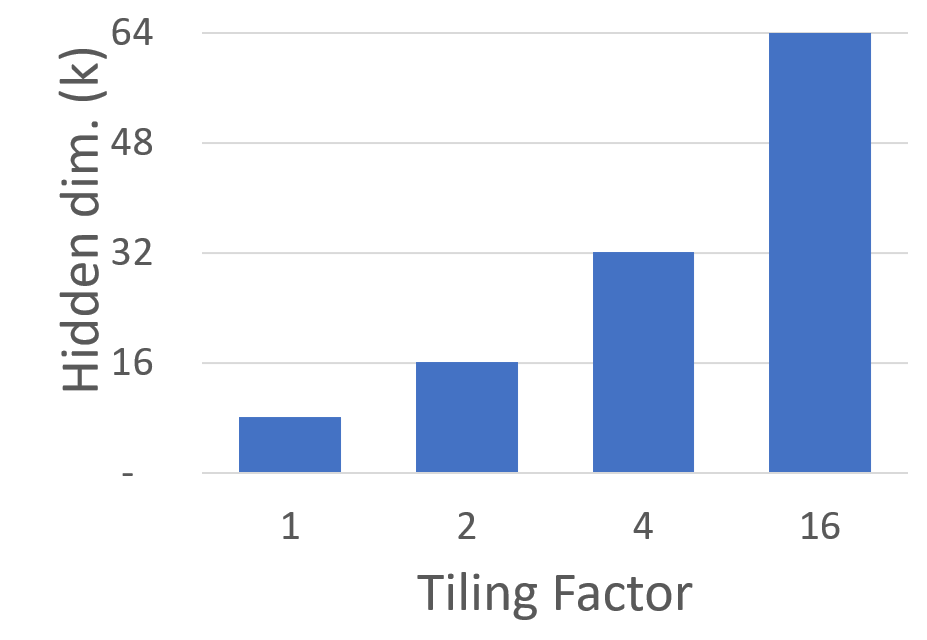}
        \captionof{figure}{\small{Max hidden dim. with different tiling factors.}}
        \label{fig:memory-centric-tiling}
    \end{subfigure}
    \hfill
    \begin{subfigure}[b]{0.19\textwidth}
        \includegraphics[width=\textwidth]{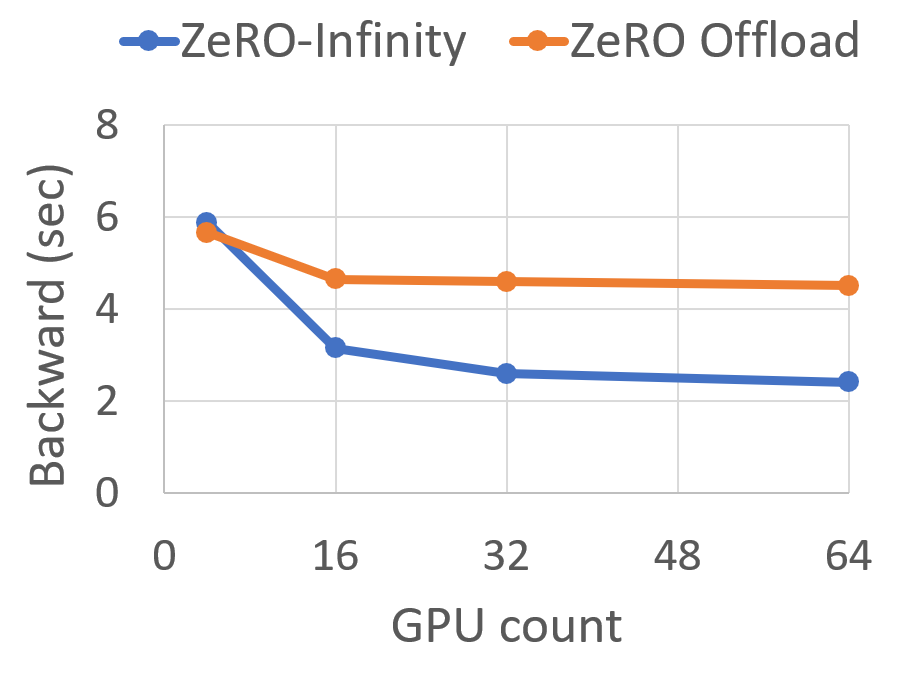}
        \captionof{figure}{\small{\name vs ZeRO Offload}}
        \label{fig:zinf-v-z2}
    \end{subfigure}
    \hfill
    \begin{subfigure}[b]{0.19\textwidth}
        \includegraphics[width=\textwidth]{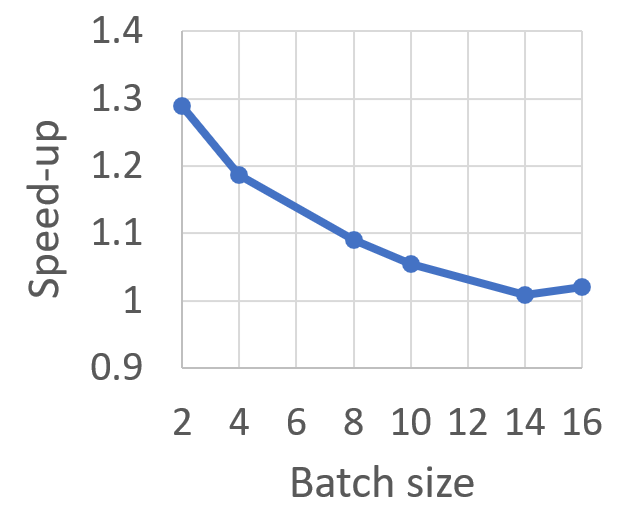}
        \captionof{figure}{\small{Speedup from communication overlap.}}
        \label{fig:prefetch-perf}
    \end{subfigure}
    \hfill
    \begin{subfigure}[b]{0.19\textwidth}
        \includegraphics[width=\textwidth]{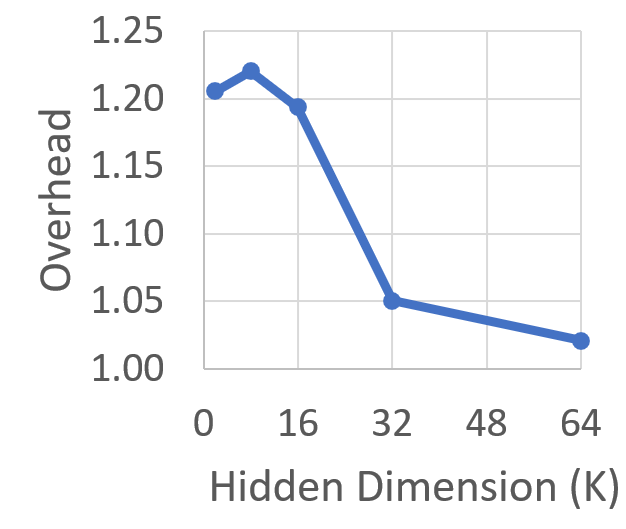}
        \captionof{figure}{\small{Overhead of offloading activation chkpt to CPU.}}
        \label{fig:activation-ckpt-perf}
    \end{subfigure}
    \caption{\small{Impact of system features on model scale and performance.}}
    \label{fig:bigperftable}
\end{figure*}

\subsection{Democratizing Large Model Training}
\Cref{fig:single_node_perf} shows performance of training 10B to 1T models on a single node (16 GPUs) with \name without any model parallelism. 
With models up to 100 billion parameters, \name achieves excellent performance of over 40 TFlops/GPU, making it possible to fine-tuning models such as GPT-3 with just a single DGX-2 box. In contrast, 3D parallelism is unable to scale to models with over 20 billion parameters. 

These results demonstrate two aspects of \name: i) Accessibility to fine-tuning large models with up to a trillion parameters on a single NVIDIA DGX-2 node, empowering users without access to large GPU clusters. ii) Ease-of-Use: Models at this scale can be trained using \name without combining model or pipeline parallelism, or requiring model code refactoring, making it easy for data scientists to scale up their models. 
    
\subsection{Impact of System Features on Model Scale}
We show the impact of different device placement strategies on model scale and impact of  memory-centric tiling (\cref{subsec:memory-centric-tiling}) on maximum hidden size using a single DGX-2 system (16 GPUs).

\begin{table}[tbh!]
  \small
  \centering
  \begin{tabular}{|c|c|c|}
    \hline
    Name & \begin{tabular}[c]{@{}c@{}} Optimizer + Grad \\ (devices/partitioned) \end{tabular} & \begin{tabular}[c]{@{}c@{}} Parameters \\ (devices/partitioned)\end{tabular}  \\ \hline
    Data parallel & [GPU] / \mycrossout & [GPU] / \mycrossout \\ \hline
    ZeRO 2         & [GPU] / \mycheckmark  & [GPU] / \mycrossout   \\\hline
    ZeRO-Offload   & [CPU,GPU] / \mycheckmark  & [GPU] / \mycrossout   \\\hline
    3D Parallelism & [GPU] / \mycheckmark  & [GPU] / \mycheckmark   \\\hline
    ZeRO 3         & [GPU] / \mycheckmark  & [GPU] / \mycheckmark   \\\hline
    ZeRO-Inf-CPU   & [CPU, GPU] / \mycheckmark  & [CPU,GPU] / \mycheckmark   \\\hline
    ZeRO-Inf-NVMe  & [NVMe,CPU,GPU] / \mycheckmark & [NVMe,CPU,GPU] / \mycheckmark  \\ \hline
\end{tabular}
  \caption{Device placement options and partitioning strategies for optimizer, gradient, and parameter states.}
\label{tab:zero-inf-configs}
\end{table}
\textbf{Maximum model size} \Cref{fig:max-model-size} shows the effect of different device placement and partitioning strategies (see \cref{tab:zero-inf-configs}) on maximum model size. By using data parallelism alone we're limited to only  1.4B parameters, due to limited GPU memory and significant model state redundancies. As we introduce optimizer/gradient partitioning and offloading to CPU with ZeRO-2 and ZeRO-Offload, we are able to scale up 9x to 13B parameters on a single node. Partitioning and offloading parameter states to CPU in \name allows us to almost reach 100B parameters. However, the final major jump in scale comes from offloading model states to NVMe which finally gets us to 1T parameters, resulting in a 700x increase in model size relative to data parallelism alone.




\textbf{Maximum Hidden Size} We evaluate the impact of memory-centric tiling in enabling large hidden sizes in the presence of memory fragmentation. We train a single layer transformer model with different hidden sizes and tiling factors to identify the largest hidden size that can be trained with and without tiling. To keep memory fragmentation consistent across all the experiments, we pre fragment the total GPU memory into 2 GB contiguous chunks so that all memory allocation requests larger than 2GB will fail.

 \Cref{fig:memory-centric-tiling} shows the largest hidden size that can be trained without memory-centric tiling is 8K, while we can even train a massive hidden size of 64K using memory-centric tiling factor of 16. With memory-centric tiling, \name greatly simplifies DL system stack by avoiding the need for model parallelism, making it easy for data scientists to train with large hidden sizes.



\subsection{Impact of System Features on Performance}
We evaluate the effects of the infinity offload engine (\cref{sec:overview}), bandwidth-centric partitioning (\cref{subsec:bandwidth-centric-partitioning}),  overlap-centric design (\cref{subsec:overlap}), and activation checkpoint offloading (\cref{sec:quantify_ait}) on training speed.

\textbf{\name vs \zoff} \Cref{fig:zinf-v-z2} shows the impact of offloading gradients to CPU memory with \name vs \zoff on the back propagation time of an 8B parameter model. \name leverages the aggregate PCIe bandwidth across GPUs to offload the gradients, resulting in a speedup of nearly 2x at 64 GPUs compared to \zoff which is limited by single PCIe bandwidth.

\textbf{Prefetching and Overlapping}
\Cref{fig:prefetch-perf} shows the relative throughput difference with communication overlapping and prefetching turned on and off for an 8B parameter model with 64 GPUs. The figure shows that prefetching and overlapping are crucial to achieving good performance at small batch sizes per GPU, while its impact diminishes at large batch sizes.

\textbf{Activation checkpoint offload}
\Cref{fig:activation-ckpt-perf} shows that CPU offloading of activation checkpoints in \name reduces the training throughput by up to 1.2x for small hidden sizes, but for hidden sizes 32K and 64K, the impact is minimal, demonstrating that it is possible to offload activation checkpoints to CPU memory without impacting efficiency for large hidden sizes.


\section{Conclusion \& Future Implications}
\label{sec:implication}
\begin{table}[tbh!]
\begin{tabular}{|c|c|c|c|}
\hline
                          & V100 & 10x  & 100x  \\ \hline
Total devices                    & 512  & 512  & 512  \\ \hline
Achievable peak (pflops/device)  & 0.07 & 0.70 & 7.00  \\ \hline
\begin{tabular}[c]{@{}c@{}}Slow memory bw requirement \\ (GB/s per device)\end{tabular} & 3.0  & 30.0 & 300.0 \\ \hline
Slow memory aggregate bw (TB/s)  & 1.5  & 15.0 & 150.0 \\ \hline
GPU-to-GPU bw (GB/s)           & 70.0 & 700.0 & 7000.0 \\ \hline

\end{tabular}
\caption{\small{Bandwidth (bw) requirements for \name to remain efficient on a cluster of 512 accelerator devices with 10x and 100x more achievable compute than NVIDIA V100 GPUs.}}
\label{tab:future_hardware}
\vspace{-5mm}
\end{table}

In this paper, we presented \name, a novel heterogeneous system technology that leverages GPU, CPU, and NVMe memory to allow for unprecedented model scale that is accessible and easy to use while achieving excellent efficiency. It offers a paradigm shift in how we think about memory for large model training. It is no longer necessary to fit DL training on ultra-fast but expensive and limited memory like HBM2. \name demonstrates that it is possible to transcend the GPU memory wall by leveraging cheap and slow, but massive, CPU or NVMe memory in parallel across multiple devices to achieve the aggregate bandwidth necessary for efficient training on current generation of GPU clusters. 

As we look into the future, the GPUs and other accelerators will become more powerful, and this aggregate bandwidth required for efficient training will also increase. \cref{tab:future_hardware} shows that even when the compute of the accelerators increases by 10x compared to the NVIDIA V100 GPUs, on a cluster with 512 of them, \name only requires a bandwidth of 30~GB/s between each accelerator and the slow memory to remain efficient.
In fact, this is already possible with today's technology by connecting accelerators to slow memory via NVLink~\cite{nvlink}. For example, the Summit Supercomputer launched in 2018~\cite{summit} connects NVIDIA V100 GPUs with the CPU memory at 40GB/s per GPU. 

It is clear that with \name, accelerator device memory is no longer a limitation on model scale or training efficiency. However, training models with tens or hundreds of trillions of parameters in a reasonable time still requires massive leaps in compute, and running efficiently on these future devices requires a proportional leap in device-to-device bandwidth (\cref{tab:future_hardware}). 

We hope that, with device memory no longer a limitation, \name will inspire a more compute and device-device bandwidth focused innovations of ultra-powerful accelerator devices and super-computing clusters in the future to support the next 1000x growth in model scale and the advancements that they can offer.
\section*{Acknowledgement}
We thank Elton Zheng, Reza Yazdani Aminabadi, Arash Ashari for their help on improving various components of the code, and Cheng Li for her help in proof reading the paper. We thank Andrey Proskurin, Gopi Kumar, Junhua Wang, Mikhail Parakhin, and Rangan Majumder for their continuous support.

\bibliographystyle{unsrt}
\interlinepenalty=10000
\bibliography{references}

\clearpage

\onecolumn

\appendix
\section{Appendix}
\label{sec:appendix}

\begin{table*}[th!] \small \begin{tabular}{|l|l|l|l|l|l|l|l|} \hline \multicolumn{8}{|c|}{Figure~\ref{fig:bigperftable}(a)} \\ \hline Model size & Number of GPUs & MP & Layers & Hidden Size & Attention head & Batch size & Total batch size \\ \hline 1.4B & 16 & 1 & 40 & 1536 & 16 & 1 & 16 \\ \hline 10B & 16 & 1 & 50 & 4096 & 16 & 1 & 16 \\ \hline 13B & 16 & 1 & 64 & 4096 & 16 & 1 & 16 \\ \hline 20B (ZeRO-3) & 16 & 1 & 98 & 4096 & 32 & 1 & 16 \\ \hline 20B (3D Par.) & 16 & 4 & 98 & 4096 & 32 & 1 & 16 \\ \hline 70B & 16 & 1 & 125 & 8192 & 32 & 1 & 16 \\ \hline 1000B & 16 & 4 & 128 & 25600 & 256 & 5 & 20 \\ \hline \end{tabular} \caption{Model configurations for Figure~\ref{fig:bigperftable}(a)} \vspace{-5mm} \end{table*}
\begin{table*}[h] \small \begin{tabular}{|l|l|l|l|l|l|l|l|} \hline \multicolumn{8}{|c|}{Figure~\ref{fig:bigperftable}(b)} \\ \hline Hidden size & Number of GPUs & MP & Layers & Model size & Attention head & Batch size & Total batch size \\ \hline 8192 & 16 & 1 & 1 & 900M & 16 & 1 & 16 \\ \hline 16384 & 16 & 1 & 1 & 3B & 16 & 1 & 16 \\ \hline 32768 & 16 & 1 & 1 & 13B & 16 & 1 & 16 \\ \hline 65536 & 16 & 1 & 1 & 50B & 32 & 1 & 16 \\ \hline \end{tabular} \caption{Model configurations for Figure~\ref{fig:bigperftable}(b)} \vspace{-5mm} \end{table*}
\begin{table*}[h] \small \begin{tabular}{|l|l|l|l|l|l|l|l|} \hline \multicolumn{8}{|c|}{Figure~\ref{fig:bigperftable}(c)} \\ \hline Number of GPUs & Hidden size & MP & Layers & Model size & Attention head & Batch size & Total batch size \\ \hline [4,16,32,64] & 8192 & 1 & 10 & 8B & 16 & 2 & [8,32,64,128] \\ \hline \end{tabular} \caption{Model configurations for Figure~\ref{fig:bigperftable}(c)} \vspace{-5mm}\end{table*}
\begin{table*}[h] \small \begin{tabular}{|l|l|l|l|l|l|l|l|} \hline \multicolumn{8}{|c|}{Figure~\ref{fig:bigperftable}(d)} \\ \hline Batch size & Number of GPUs & Hidden size & MP & Layers & Model size & Attention head & Total batch size \\ \hline [2,4,8,10,14,16] & 64 & 8192 & 1 & 10 & 8B & 16 & [128,256,512,640,896,1024] \\ \hline \end{tabular} \caption{Model configurations for Figure~\ref{fig:bigperftable}(d)} \vspace{-5mm}\end{table*}
\begin{table*}[h!] \small \begin{tabular}{|l|l|l|l|l|l|l|l|l|} \hline \multicolumn{9}{|c|}{Figure~\ref{fig:bigperftable}(e)} \\ \hline Hidden size & Number of GPUs & Opt Device & MP & Layers & Model size & Attention head & Batch size & Total batch size \\ \hline 2048 & 32 & CPU & 1 & 5 & 275M & 16 & 4 & 128 \\ \hline 8192 & 32 & CPU & 1 & 5 & 4B & 16 & 4 & 128 \\ \hline 16384 & 32 & CPU & 1 & 5 & 16B & 16 & 4 & 128 \\ \hline 32768 & 32 & CPU & 1 & 5 & 64B & 16 & 4 & 128 \\ \hline 65536 & 64 & NVMe & 1 & 5 & 260B & 16 & 4 & 128 \\ \hline \end{tabular} \caption{Model configurations for Figure~\ref{fig:bigperftable}(e)} \end{table*} 


\end{document}